\documentstyle[aps,preprint]{revtex}
\begin{document}
\draft

\title{ Distribution of "level velocities"\\
in quasi 1D disordered or chaotic systems with localization.}
\author{ Yan V. Fyodorov}
\address{ Fachbereich Physik, Universit\"{a}t-GH Essen,
D-45117 Essen, Germany \\ and \\ Petersburg Institute of Nuclear
Physics, Gatchina 188350, Russia} \date{May 25, 1994}

\maketitle

\begin{abstract}
The explicit analytical expression for the distribution
 function of parametric derivatives of energy levels
("level velocities") with respect to a random change of
scattering potential is derived for the chaotic quantum
systems belonging to the quasi 1D universality class
(quantum kicked rotator, "domino" billiard, disordered
wire, etc.).  \end{abstract}


It is generally accepted nowadays that the problem of a
quantum particle moving in a random potential
first addressed in the context of Anderson
localization has much in common with such problems in the
domain of Quantum Chaos as "quantum kicked rotator"
\cite{Izrrev}, the ionization of Rydberg atoms by a
microwave radiation \cite{CCGS} and chains of quantum
billiards (the "domino billiard" \cite{Dombil}). This analogy
first suggested in \cite{Fish} proved to be very fruitful
for understanding of the phenomenon  of the so-called
"dynamical localization"  \cite{Izrrev}.

More recently it was found that there exists a convenient
mathematical framework -- the ensemble of Random Banded
Matrices (RBM), see \cite{RBM,FMRev} and references therein --
generalizing the classical
Gaussian Ensembles of Random Matrices and allowing for a uniform
description of the typical features common to the
abovementioned systems. Investigation of the RBM
ensemble helped to reveal a number of universal scaling
relations characterizing statistics of energy levels and
eigenfunctions of all these systems\cite{Izrrev,RBM,FMlett}.
Another important feature is that the stochastic RBM model
can be mapped onto a regular field-theoretical model --
a so-called nonlinear graded $\sigma-$model --
allowing in some cases for an exact
analytical treatment and so providing one with a powerful
tool of research\cite{FMlett,FMRev}. This nonlinear
$\sigma-$model turns out to be identical to that derived
earlier by Efetov and Larkin in the course of
study of the Anderson localization in disordered wires
\cite{EL}.

All these facts suggest to introduce a notion of a "quasi
$1D$ universality class" of disordered and chaotic systems.
All statistical properties of systems belonging to this
class are dependent on the only scaling parameter: the
ratio $x=L/\xi$ between the sample length $L$ and the
localization length $\xi$. The explicit form of the scaling
function was derived analytically for the so-called inverse
participation ratio measuring the extent of eigenfunctions.
For other quantities characterizing eigenfunction
statistics analytical results are available in two limiting
cases $x\gg 1$ ($x\ll 1$) corresponding to the complete
localization (delocalization) of eigenfunctions
\cite{FMlett,FMRev}. As to the statistics of energy levels,
only heuristic expressions deduced from the numerical data were
available so far \cite{Izrrev}.

Quite recently an interesting new developement in study of
weakly disordered metallic systems and their chaotic
counterparts has been made in a set of works by the MIT group
\cite{Alt}. Developing earlier ideas from the papers
\cite{PNV} the authors of \cite{Alt} studied the energy
level motion as a function of some external tunable
parameter $\alpha$. Physically the role of such a parameter
can be played by  e.g an external magnetic field, the
strength of a scattering potential for disordered metal, a
form of confining potential for quantum billiards, or any
other appropriate parameter which the system Hamiltonian is
dependent on. A high degree of universality in a "level
response" of a generic chaotic system to an external
perturbation has been revealed. It was found that a set of
"level velocities" (LV) $v_{n}(\alpha)\equiv \partial
E_{n}/\partial \alpha$ , with index $n$ labeling different
energy levels, can be characterized after a proper
normalization by universal correlation functions $\langle
v_{n}(\alpha)v_{n'}(\alpha')\rangle$ whose form is dependent only
on symmetries of the unperturbed Hamiltonian and those of
perturbations. Related quantities characterizing energy
level response, such as the distribution of "level
curvatures" $K_{n}\equiv \partial^{2}v_{n}/\partial
\alpha^{2}$ and that of "avoided crossings" (local minima
of adjacent level spacings) were studied in the
papers \cite{PNV,Del}. Let us also mention an intimate connection
between the "level response" characteristics
and the system conductance if the role of the perturbation
is played by the Aharonov-Bohm (AB) magnetic flux
, see the detailed discussion of the
issue in \cite{Akm}.

The most of analytic work on the "level response"
characteristics done so far have made use of the analogy
"Quantum Chaotic Systems -- Random Matrix Ensembles" which
is now a commonly accepted principle in the domain of Quantum
Chaos\cite{RM}. However, a simulation of
chaotic (disordered) systems by the classical Gaussian
Random Matrix Ensembles (or equivalent models) precludes
effects of Anderson localization from being taken into
account. On the other hand, these effects should
considerably modify all the results when one deals with
systems belonging to the "quasi 1D universality class"
introduced above. This indeed was found to be the case in
the numerical study of the curvature distribution
for the {\it periodic} RBM simulating a
disordered ring threaded by the AB flux
\cite{RBMper}.

In this letter I derive, for the first time, the distribution
of the "level velocities" for the systems belonging to the quasi 1D
universality class in the most interesting
limit of infinite sample length when the role of
the localization effects is expected to be maximal.
As the particular model for unperturbed system I use the
ensemble of {\it nonperiodic} Hermitean RBM simulating a
quasi $1D$ system ( an isolated piece of wire, an irregular
billiard chain, etc.) subject to a magnetic field.
The class of perturbations considered corresponds to a
 slight random change of scattering potential within the
wire. For this reason obtained results
are not valid for systems with {\it
 periodic} geometry and AB flux playing a role of
 perturbation. The latter case requires therefore for a
separate investigation.

Let us consider an unperturbed chaotic or disordered isolated
quasi 1D system of finite size $L$ having $N\propto L$ energy levels
$E_{n},\,\, n=1,2,...,N$ and described by a Hamiltonian ${\cal H}$ whose
statistical properties are adequately simulated by those of the RBM ensemble.
Let us study the "level response" to a perturbation $\delta {\cal H}=\alpha
{\cal V}$, with $\alpha$ being a small parameter.
For this purpose it is convenient to introduce the resolvent operator
$R^{\pm}_{\alpha}(E)=[E-{\cal H}-\alpha {\cal V}\pm i\epsilon]^{-1}$, with
$\epsilon$ being a positive infinitesemal. Then one has the following
selfevident identity:
\begin{equation}
\lim_{\epsilon\to 0} \epsilon Tr R_{0}^{+}(E) Tr R_{\alpha}^{-}(E)=
\lim_{\epsilon\to 0}\sum_{n,m}\frac{\epsilon}{[E-E_{n}(0)+i\epsilon]
[E-E_{m}(\alpha)-i\epsilon]}\label{1}\end{equation}

It is well known that for the chaotic quasi 1D system of any {\it finite}
size the spectrum consists of the set of {\it nondegenerate} levels, the
probability ${\cal P}(\delta E)$
for two adjacent levels to be separated by a gap $\delta E$
tending to zero when $\delta E \to 0$ (see, e.g. the so-called "Izrailev
distribution" \cite{Izrrev} and the recent analytical results
by Kolokolov on the related subject \cite{Kol}). Let us now impose
the following requirements: (i) the limiting procedure
in eq.(\ref{1}) to be performed {\it prior} to the
thermodynamic limit $L\to \infty$ and (ii) the parameter
$\alpha\propto \epsilon \rightarrow 0$ when performing the limiting procedure.
Then the only nonvanishing contribution to the double sum in eq.(\ref{1})
is given by pairs of coinciding level indices $n=m$.
Introducing the notations $\mu=2\epsilon/\alpha$ and $v_{n}=\partial E_{n}/
\partial\alpha\mid_{\alpha=0}$ one obtains:
\begin{equation}\label{2}
K(\mu)=\lim_{\epsilon\to 0}\epsilon Tr R_{0}^{+}(E)Tr R^{-}_{2\epsilon/\mu}(E)
=\pi \sum_{n}\frac{\mu^{2}+i\mu v_{n}}{\mu^{2}+v_{n}^{2}}\delta(E-E_{n})
\end{equation}

Performing now formally the averaging over the ensemble of Hamiltonians
${\cal H}$ (denoted by the angular brakets $\langle ... \rangle$)
and introducing the mean level density $\rho(E)$ one immediately
finds the relation between $K(\mu)$ and the distribution ${\cal P}(v)$
of the "level velocities":
\begin{equation}\label{3}
\displaystyle{\frac{1}{N\rho}}Re\langle K(\mu)\rangle = \pi \mu^{2}
\int_{-\infty}^{\infty}\frac{{\cal P}(v)}{v^{2}+\mu^2}dv\equiv \pi\mu
\int_{0}^{\infty}e^{-\mu k}dk \int_{-\infty}^{\infty} {\cal P}(v)\cos{kv}dv
\end{equation}

The relations presented above are actually valid for
an arbitrary chaotic or disordered system. Let us now specify the Hamiltonian
${\cal H}$ as being a $N\times N$ Hermitean random matrix with independent
gaussian distributed entries ${\cal H}(i,j)$ with zero mean
and the variance $\langle {\cal H}^{*}(i,j){\cal H}(i,j)\rangle=\displaystyle{
a\left(\frac{\mid i-j\mid}{b}\right)}[1+\delta_{ij}]$.
Here the function $a(r)$ is
assumed to vanish at $r\to \infty$ at least exponentially fast, the parameter
$b$ (assumed to be large: $b\gg 1$)
 defining therefore the effective bandwidth of RBM.

In order to perform the ensemble average I employ the
Efetov's supersymmetric approach \cite{EL}. The detailed
exposition of the method as applied to the RBM ensemble can be found in
\cite{FMRev} and is not repeated here. After performing all the
necessary steps the problem is mapped onto a $1D$ nonlinear
$\sigma-$ model with the action $S[Q]=S_{0}[Q]+\delta S[Q]$, where
\begin{equation}\label{4}
\begin{array}{l} S_{0}[Q]=\displaystyle{\frac{\gamma}{4}}
\sum_{i=1}^{N}Str(Q_{i}-Q_{i+1})^{2}
+i\pi\rho\epsilon \sum_{i=1}^{N}Str Q_{i}\Lambda \\
\delta S[Q]=\sum_{k=1}^{\infty}\displaystyle{\frac{1}{k}
\left(-\frac{2i\pi\rho\epsilon}{\mu}\right)^k\sum_{i_{1},..,i_{k}}}
{\cal V}(i_{1},i_{2}){\cal V}(i_{2},i_{3}) ...
{\cal V}(i_{k},i_{1}) Str \prod_{m=1}^{m=k}Q_{i_{m}}
\displaystyle{\frac{1-\Lambda}{2}}.\end{array}\end{equation}
Here $Str$ stands for the supertrace \cite{EL}, the
$4\times 4$ matrices $Q_{i}$
belong to the graded coset space $U(1,1/2)/U(1,1)\times U(1,1)$
\cite{EL,FMRev} and
$\Lambda=diag (1,1,-1,-1)$.

The nonlinear $\sigma-$ model with the action $S_{0}[Q]$ describes
the statistical properties of the unperturbed quasi $1D$ system and
it was intensively studied in \cite{FMlett,FMRev}. The main parameter is
the coupling constant $\gamma$ expressed in terms of the RBM parameters
as follows: $\gamma= (\pi\rho)^{2}\sum_{r}a(r)r^{2}\propto b^{2}$
\cite{FMlett,FMRev}. It defines
the only characteristic spatial length scale due to disorder: the localization
length $\xi\propto \gamma$. On the more formal level it plays the role of the
correlation length of the matrix field $Q_{i}$. That means
 that the matrices $Q_{i}$ and $Q_{j}$ can be considered as equal to each other
as long as $\mid i-j \mid \ll \gamma$. Let us now make a natural assumption
that the spatial structure of the perturbation ${\cal V}$
is of the same type as that of the unperturbed Hamiltonian ${\cal H}$, i.e.
${\cal V}(i,j)$ vanishes sufficiently fast as long as $\mid i-j \mid\gg b$.
We will call such a perturbation ${\cal V}$ the generic one. Then in view of
the relation $1\ll b\ll b^{2}\propto\gamma$ one can put $Q_{i_{1}}=
Q_{i_{2}}=...=Q_{i_{k}}$ in the expression for $\delta S[Q]$.

{}From the previous experience \cite{FMlett,FMRev} one can anticipate
that the main contribution to the integrals over the matrix field $Q_{i}$
is coming from the asymptotic domain $Q_{i}\sim 1/(\epsilon\gamma)$ as long
as $\epsilon\rightarrow 0$.
As it will be clear afterwards, the essential
values of the parameter $\mu$ are of the order of
$\mu\sim max\{\gamma^{-1/2}, N^{-1/2}\}$. For a generic random perturbation
one notices that
${\cal V}_{0}\equiv ({\cal V}^{2})(i,i)=\sum_{m}\mid{\cal V}(i,m)\mid ^{2}$
is a deterministic (selfaveraging) quantity of the
order of unity, whereas ${\cal V}(i,i)$ is a random quantity
with zero mean and variance of the order of
$\langle{\cal V}(i,i)^{2}\rangle\sim {\cal V}_{0}/b\ll {\cal V}_{0}$. Combining
all these estimates together one arrives at the final form for the
effective action of the nonlinear $\sigma-$ model:
\begin{equation}\label{6}
S[Q]=S_{0}[Q]-\frac{1}{2}\left(\frac{\pi\rho\epsilon}{\mu}\right)^{2}
{\cal V}_{0} Str\sum_{i}Q_{i}(1-\Lambda)Q_{i}(1-\Lambda).
\end{equation}

In view of the local-in-space structure of the last term in eq.(\ref{6})
the corresponding integral over the matrices $Q_{i}$
can be performed by the same transfer-matrix method used earlier
\cite{FMlett,FMRev} with minor modifications. As the result one finds
$\frac{1}{\rho N}\langle K(\mu)\rangle =\frac{1}{x}I(x)$ where
\begin{equation}\label{7}
I(x)=\int_{0}^{x}d\tau_{1}\int_{0}^{x-\tau_{1}}d\tau \int_{0}^{\infty}
\frac{dy}{y} Y^{(1)}(x-\tau-\tau_{1};y)Y^{(2)}(\tau,\tau_{1};y)
\end{equation}
with both functions $Y^{(1)}(\tau,y)$ and $Y^{(2)}(\tau,\tau_{1},y)$
satisfying the same differential equation:
\begin{equation}\label{8}
\frac{\partial Y}{\partial \tau}={\cal G}Y\quad;\quad {\cal G}=y^{2}
\frac{\partial^{2}}{\partial y^{2}}-\left(y+\frac{y^{2}}{4g^{2}}\right).
 \end{equation}
Here the "scaling" notations $g=\displaystyle{\frac{\mu}{\pi\rho}\left[
\frac{\gamma}{2{\cal V}_{0}}\right]^{1/2}}$ and
$x=\displaystyle{\frac{N}{2\gamma}}$ were introduced
for the sake of convenience. The equation eq.(\ref{8})
should be supplied with the "initial" conditions:
\begin{equation}\label{9}
Y^{(1)}(\tau=0;y)=1\quad,\quad Y^{(2)}(\tau=0,\tau_{1};y)=
y Y^{(1)}(\tau_{1};y),\end{equation}
the first one corresponding to the elastic reflection of the quantum particle
at the sample edges and the second one related to the details
of the transfer matrix method \cite{FMRev}.

In order to be able to deal with the function $I(x)$ efficiently it is
more convenient to consider its Laplace transform
$I_{L}(p)=\int_{0}^{\infty}e^{-px} I(x) dx$. One finds:
\begin{equation}\label{10}
I_{L}(p)=\int_{0}^{\infty}\frac{dy}{y}Y^{(1)}_{L}(p;y)Y^{(2)}_{L}(p;y)
\end{equation}
where the functions $Y_{L}^{(1),(2)}(p,y)$ satisfy the
system of two ordinary differential equations:
\begin{equation}\label{11}
{\cal G}_{p}Y_{L}^{(1)}(p;y)=-\frac{1}{y^{2}}\,\,,\,\, {\cal G}_{p}
Y_{L}^{(2)}(p;y)=-\frac{1}{y}Y_{L}^{(1)}(p;y)\quad;\,\,
{\cal G}_{p}=\frac{\partial^{2}}{\partial y^{2}}-
\left(\frac{1}{4g^{2}}+\frac{1}{y}+\frac{p}{y^{2}}\right).\end{equation}
One can write down the explicit solution to these equations noticing that
the two Whittaker functions $W_{-g,\frac{\kappa}{2}}(y/g)\equiv
\phi^{\kappa}_{1}(y)$
 and $M_{-g,\frac{\kappa}{2}}(y/g)\equiv \phi^{\kappa}_{2}(y)$
are eigenfunctions of the operator
${\cal G}_{p}$ corresponding to the eigenvalue $\lambda_{\kappa}=
-\frac{1}{4}(4p+1-\kappa^{2})$ and the corresponding Wronskian is equal to
$w_{\kappa}=-\Gamma [\kappa+1]/\Gamma [g+(\kappa+1)/2]$,
 where $\Gamma [z]$ stands for the
Euler gamma-function. Thus, one finds $Y_{L}^{(2)}(p;y)=
{\cal R}\{y Y_{L}^{(1)}(p;y)\}$ , where
\begin{equation}\label{12} Y_{L}^{(1)}
(p;y)=\frac{1}{p} \Gamma [1+g] W_{-g,1/2}(y/g)+{\cal R}
\{1-\Gamma [1+g] W_{-g,1/2}(y/g)\}\end{equation}
and the action of the operator ${\cal R}$ on any function $f(y)$
 is defined by ( $\nu=\sqrt{4p+1}$) :
\begin{equation} \label{13}
{\cal R}\{f\}=\frac{-1}{w_{\nu}}\left\{
\phi^{\nu}_{1}(y)\int_{0}^{y}\phi^{\nu}_{2}(z)f(z)\frac{dz}{z^{2}}+
\phi^{\nu}_{2}(y)\int_{y}^{\infty}\phi^{\nu}_{1}(z)f(z)\frac{dz}{z^{2}}
\right\} \end{equation}

Expressions eqs.(\ref{11}-\ref{13}) provide a formal possibility to find
 the function $I_{L}(p)$ for an arbitrary value of $p$. Actually, however,
managable expressions could be extracted only in the limiting case
$p\to 0$ physically corresponding to the system length $L$ being much larger
than the localization length $\xi$, i.e. $x\propto L/\xi \gg 1$. The
 main simplification occurs if one notices that one can
neglect the second term in the expression for $Y^{(1)}(p;y)$, eq.(\ref{12}),
provided $p\to 0$. As the result the eqs.(\ref{10})-(\ref{11})
can be presented in the form:
\begin{equation}\label{14}\begin{array}{c}
I_{L}(p\to 0)=-g\frac{\Gamma^{2}[1+g]}{p^{2}}\int_{0}^{\infty}\frac{dz}{z}
W_{-g,1/2}(z)\tilde{Y}^{(2)}(z)\\
{\cal L}\tilde{Y}^{(2)}(z)=\frac{1}{z}W_{-g,1/2}(z)\quad;\quad
 {\cal L}= \frac{\partial^{2}}{\partial z^{2}}-\left(
\frac{1}{4}+\frac{g}{z}\right)\end{array}\end{equation}

Differentiating the identity ${\cal L}W_{-g,1/2}(z)=0$ over the parameter
$g$ and using the condition $\tilde{Y}^{(2)}(z\to 0)=0$
one obtains:
\begin{equation}
\label{15}
\tilde{Y}^{(2)}(z)
=\left[\frac{\partial}{\partial g}+ \psi(1+g)\right]
W_{-g,1/2}(z)  \quad \mbox{where}\quad
\psi(z)=\frac{\partial \ln{\Gamma[z]}}{\partial z}\end{equation}

Substituting this expression into eq.(\ref{14}) and remembering the relation
between $\langle K(\mu) \rangle$ and $I(x)$ one finds:
\begin{equation}\label{16}
\frac{1}{N\rho}\langle K(\mu)\rangle \mid_{x\rightarrow \infty}=
g\left[\frac{\partial \psi(g)}{\partial g}+\frac{1}{2}g
\frac{\partial^{2}\psi(g)}{\partial g^{2}}\right] \equiv \frac{g}{2}
\int_{0}^{\infty} dk e^{-g k}\left[\frac{k/2}{\sinh{(k/2)}}\right]^{2}
 \end{equation}

Introducing now the scaled "level velocity" $v_{s}=
v\frac{1}{\pi\rho}
\left(\frac{\gamma}{2{\cal V}_{0}}\right)^{1/2}$ and comparing
the eq.(\ref{16}) with the eq.(\ref{3}) one restores the LV distribution
function ${\cal P}(v_{s})$ from its Fourier transform:
\begin{equation}\label{17}
{\cal P}(v_{s})=\int_{0}^{\infty}\frac{dk}{2\pi}\cos{kv_{s}}\left[
\frac{k/2}{\sinh{(k/2)}}\right]^{2} =\frac{\pi}{\sinh^{2}{(\pi v_{s})}}
\{\pi v_{s}\coth{(\pi v_{s})}-1\}\end{equation}
This expression gives the explicit form of the LV distribution
for the case of long quasi $1D$ disordered or chaotic system
and is the main result of the present Letter.

Let us briefly mention that in the opposite limiting case of short systems
whose length $L\ll \xi$ the effects of localization play no role and one
easily reproduces from eqs.(\ref{7}-\ref{8})
the Gaussian LV distribution typical for the chaotic
systems studied in earlier papers \cite{Alt,Del}.
solution to the eq.(7) in the domain $0\le\tau\le x\ll 1$
is given by the expression:
\begin{equation}
Y(\tau;y)=Y(\tau=0;y)exp{-\tau\left(
y+\frac{y^2}{4g^2}\right)}
\end{equation}
that immediately produces the required result when
substituted to the eqs.(\ref{7},\ref{3}).

Comparing the two limiting cases one concludes
that the "level velocities" fluctuate much stronger when eigenfunctions are
localized: (i) the probability to find values of LV  exceeding the typical
value $\langle v^{2}\rangle^{1/2}$ decays in the case of extended
states like $e^{-c\, v^{2}}$, i.e. much faster than a simple exponential
typical for localized states, see eq.(\ref{17}) and (ii)
the mean square $\langle v^{2}\rangle$ is proportional to the inverse
localization length $1/\xi\propto 1/\gamma$ when localization takes place,
i.e. is much larger than the value of the order
of inverse system size $1/L \propto 1/N$
typical for systems with extended states and the same number of
levels $N$. To this end it is interesting to note
that in the papers \cite{Alt}
the quantity $\langle v^{2}\rangle$
was called the "generalized conductance" in view of its meaning
for the AB case\cite{Akm}.
The results obtained in the present paper suggest that the level response
of a disordered system subject to a random perturbation
is rather related to the so-called "inverse participation ratio" which is
inversely proportional to the eigenfunction extent. A more detailed
discussion of this issue will be published elsewhere
\cite{FM}.

It seems to be interesting to check all this predictions by a
direct numerical simulations of the systems belonging to the quasi $1D$
universality class. In particular, for a chain of chaotic quantum billiards
( the "domino billiard"\cite{Dombil}) one should be able to observe a
substantial increase in LV fluctuations when passing from the
regular chain to an irregular one.

The author is grateful to I.Kolokolov for providing him with results
of the paper \cite{Kol} prior to the publication and to H.-J.Sommers
for valuable discussions. The financial support by the program SFB 237
"Unordnung und grosse Fluctuationen" is acknowledged with thanks.

\end{document}